\documentstyle[aasms4]{article}

\begin{document}
\title{{\large {\bf Sub-degree CMB anisotropies from
inflationary bubbles}}}
\author{Carlo Baccigalupi}

\affil{INFN and Dipartimento di Fisica, 
Via del Paradiso 12, 44100, Universit\`a di Ferrara; 
Osservatorio Astronomico di Roma, Viale del Parco Mellini, 84, 
00136 Roma}
 
\baselineskip 10pt
\vspace{2.cm} 
\begin{abstract}
It is well known that processes of first order phase transitions
may have occurred in the inflationary era. 
If one or more occurred well before the end of inflation, 
the nucleated bubbles are stretched to large scales
and the primordial power spectrum contains a scale dependent
non-Gaussian component provided by the remnants of the bubbles.
We predict the anisotropies
in the cosmic microwave background (CMB) induced by   
inflationary bubbles. We build a general analytic model for
describing a bubbly perturbation; we evolve each Fourier
mode using the linear theory of perturbations from reheating until 
decoupling; we get the CMB anisotropies by considering the bubbly
perturbation intersecting the last scattering surface.
The CMB image of an inflationary bubble is a series of
concentric isothermal rings of different color (sign of $\delta T/T$)
on the scale of the sound horizon at decoupling ($\le 1^{o}$ 
in the sky); the resulting anisotropy is
therefore strongly non-Gaussian. The mean amplitude of
$\delta T/T$ for a bubble of size $L$ follows the known estimates 
for linear perturbations,  
$\delta T/T\simeq\delta\rho /\rho\cdot (L/H^{-1})^{2}$.
In particular, bubbles with size corresponding to the seeds of
the observed large scale voids (tens of comoving Mpc) induce an 
interesting pattern of CMB anisotropies on the sub-degree angular 
scale, to be further investigated and compared with the forthcoming 
high resolution CMB maps provided by the MAP and the Planck 
experiments.
\end{abstract}


	\section{\normalsize\bf Introduction}

Inflation is a phase of accelerated expansion driven 
by very high energy quantum fields ($kT\simeq 10^{15}$ GeV),
which leaves observable cosmological traces at the present.
The perturbations we observe today in the universe are
therefore the most direct link we have to very high energy physics.
Today, this fundamental topic is under investigation from several 
points of view: the perturbations imprinted on
the cosmic microwave background (CMB) at the recombination era,
the possible traces on the cosmic background of gravitational waves,
the direct reconstruction of the three dimensional matter distribution 
traced by galaxies and their peculiar velocities in the modern 
redshift surveys and others. 

The existence of large voids in the galaxy distribution, known for 
a long time (\cite{AV}), is now being established quite firmely.
Recently, several almost spherical voids with mean diameter 
of $40{\it h}^{-1}\ {\rm Mpc}$ have been detected and catalogued in 
the SSRS2 survey (\cite{EPD1}); the surrounding walls 
have a thickness smaller than 10${\it h}^{-1}\ {\rm Mpc}$; the largest
voids reach the spatial limits of the survey and their fraction of
volume is about $40\%$; if smaller voids are also considered 
the voids volume fraction has a lower limit of $75\%$.
These amazing results are confirmed by the analysis on IRAS 
(\cite{EPD2}); these papers contain very
impressive pictures reproducing the large voids
in a three dimensional view. Although rare isolated
faint galaxies are found inside the voids, preliminary analysis
of the peculiar velocities of the walls galaxies (\cite{DFWGHS}) 
indicate the almost complete absence of matter (baryonic and dark) 
in the central cavities. The matter is therefore confined
between the voids, and this surely affects
the cluster distribution, that recently has been 
found more correlated than expected (see e.g. \cite{CLUSTER} and
references therein); also, the matter distribution on 
$\sim 100h^{-1}$ Mpc is revealing an evidence of extra 
power (\cite{100MPC}).
The observed inhomogeneities constrain
the minimal inflationary scenario (slow rolling and purely
Gaussian fluctuations, see \cite{P}) and the consequences
upon the viable parameter space are currently under 
investigation (\cite{G} and references therein).
Topological defects provided mainly by 
{\it post}-inflationary phase transitions (\cite{TD} and references 
therein), have been proposed to be the seeds for the observed 
inhomogeneities, and their CMB imprints have been predicted 
(\cite{DP}); however, recent results revealed some power 
deficit both on the CMB and on the perturbations spectrum (\cite{BII}).
A distinctive feature of these models is a
non-Gaussian perturbation spectrum; non-Gaussianity means
that the phases of the components of the perturbations
are correlated in some way.
Getting ready to detect eventual non-Gaussianity in the
CMB high resolution maps provided by the next experiments is
a non-trivial theoretical challenge that is currently under
development (\cite{NG} and references therein). 

This paper aims at predicting the CMB
anisotropies generated by models of first order inflation in which 
the perturbation spectrum is in general the sum
of the Gaussian slow rolling fluctuations
and the scale dependent non-Gaussian perturbations from
the remnants of nucleated bubbles. First order
phase transitions in the early universe are predicted
by many inflationary theories (see \cite{FOI}), and one of the most 
interesting ideas introduced
in inflationary cosmology in recent years is that in general
the phase transition occurs {\it during} the inflationary slow
rolling (\cite{OA},\cite{ABKOR}). 
Like in ordinary first order inflation, the nucleated bubbles grow
as the causal horizon until the collision with other bubbles
stops the overcomoving growth; however in these models the
true vacuum energy is not zero, thus
the bubbles may be in the linear regime and, most important, the
amount of $e-$foldings between the phase transition and
the end of inflation allows the bubbles to 
grow comovingly. This phenomenology has very interesting
consequences for what concerns the gravitational
radiation produced by the colliding bubbles:
the characteristic peak frequency, exponentially
redshifted by inflation, may fall in a detectable range 
(\cite{BAFO}). At reheating, 
the energy stored in the bubble wall is converted in
matter and radiation, leaving a bubbly trace in the cosmic density.
In the case in which a phase transition
occurred at about 50 $e$-foldings before the end of inflation,
the remnants of the bubbles have comoving sizes of tens of Mpc;
they could be the seeds of the large voids
observed today, a possibility firstly suggested in
\cite{LA} and then further investigated 
(see e.g. \cite{ABKOR}, \cite{OBAM}).

In a previous work (\cite{BAO}) we computed the CMB 
anisotropies induced by non-linear voids and we obtained from
COBE data an upper limit of about $100{\it h}^{-1}\ {\rm Mpc}$ 
to the present radius of a 
non-linear void sitting on the last scattering surface (LSS); in
\cite{OBAM} we considered mainly the time evolution of 
spherical bubbles in the matter dominated era, as the outcome of 
linear ones at decoupling and also we obtained (again from COBE) 
an upper limit of about $100{\it h}^{-1}\ {\rm Mpc}$ to the 
present radius of linear bubbles on the LSS. 
Here we compute the detailed CMB anisotropies generated by 
an inflationary bubble.
We build an analytic model for
describing a general linear 
bubbly perturbation; 
we evolve each Fourier
mode from reheating to decoupling using the theory of
linear perturbations (see \cite{HS} and references therein). 
Finally we get the CMB anisotropies by
considering the intersection between the bubble and the LSS. 
We shall focus the analysis
on bubble comoving sizes of tens of Mpc at decoupling
(angular scale below $1^{o}$), since they correspond
to the observed structures and voids recently
catalogued (\cite{EPD1},1997).

In \cite{ABOII} we consider a realistic CMB fluctuation
field produced by a primordial power spectrum containing
ordinary Gaussian fluctuations and a non-Gaussian bubbly
component, as predicted by the first order inflation models
mentioned above; the volume fraction 
($\simeq 50\%$ at the present)
occupied by the bubbles allow them to significatively affect the 
large scale structure: the impact of this model on the CMB angular 
power spectrum is predicted.

Our computations shall be compared with the high resolution CMB
maps provided by the next MAP and Planck experiments, that will
provide the whole sky CMB anisotropy maps down to
$\delta T/T\simeq 10^{-6}$ and with angular resolution
of 10$^{'}$.

The paper is organized as follows: section II is dedicated to the
analytical construction of the
perturbation field; in Section III we apply the linear perturbation
theory and evolve the bubbly perturbation from reheating until
decoupling focusing on the corresponding perturbation in the photon
baryon plasma; in Section IV we compute the
bubbly CMB anisotropies; finally, Section V contains
the conclusions.

\section{A general analytic model} 

In principle, it is possible to calculate the exact bubble shape
by considering a specific first order inflationary model and the fields
involved in the process. However, since a bubbly density
perturbation is specified by few parameters, we choose here
to parametrize the inflationary bubble directely
at reheating. Bubbles from a phase transition occurred early during
inflation are stretched comovingly to large scales 
by the remaining inflation (\cite{OA},\cite{ABKOR}). 
At reheating they leave a perturbation in the
cosmic density.
A compensated spherical bubble is described by its central
density contrast $\delta$, comoving radius $R$ and thickness $\sigma$ 
of the compensating shell:
\begin{equation}
\delta =\left|{\delta\rho\over\rho}\right|_{r=0}\ \ ,\ \ R \ \ ,\ \
\sigma={\Delta R_{shell}\over R}\ \ .
\label{model}
\end{equation}
Defining a radial coordinate $x=r/R$ in the comoving gauge,
we model naturally two zones in a bubble, the
central underdensity and the compensating shell; the width of 
the latter is divided in inner ($\sigma_{i}$) and outer 
($\sigma_{o}$), with 
$\sigma_{i}+\sigma_{o}=\sigma$. We model the bubble as follows:
\begin{eqnarray}
{\delta\rho\over\rho} &=& -\delta\ \ (x\le 1-\sigma_{i})\ \ ,
\nonumber\\ 
{\delta\rho\over\rho} &=& A+B\sin\left[{\pi\over\sigma_{i}}
\left(x+{\sigma_{i}\over 2}-1\right)\right]
\ \ (1-\sigma_{i}\le x\le 1)\ \ ,  \label{dc}\\
{\delta\rho\over\rho} &=& C+D\sin\left[{\pi\over\sigma_{o}}
\left(x+{\sigma_{o}\over 2}-1\right)\right]
\ \ (1\le x\le 1+\sigma_{o})\ \ .\nonumber
\end{eqnarray}
The constants are determined by the requests of continuity and
compensation:
\begin{eqnarray}
D &=& C=B-{\delta\over 2}=A+{\delta\over 2}=\nonumber\\
=&\delta &{{1\over 3}\left(1-\sigma_{i}\right)^{3}+
{1\over 6}\left[1-\left(1-\sigma_{i}\right)^{3}\right]-
{\sigma_{i}^{2}\over\pi^{2}}\left(2-\sigma_{i}\right)\over
{1\over 3}\left[1-\left(1-\sigma_{i}\right)^{3}\right]+
{2\sigma_{i}^{2}\over\pi^{2}}\left(2-\sigma_{i}\right)+
{1\over 3}\left[\left(1+\sigma_{o}\right)^{3}-1\right]-
{2\sigma_{o}^{2}\over\pi^{2}}\left(2+\sigma_{o}\right)}\ \ .
\label{abcd}
\end{eqnarray}
The Fourier transform 
of the above density contrast in a volume $V$ is
\begin{eqnarray}
\left({\delta\rho\over\rho}\right)_{k}=
\int{\delta\rho\over\rho}(r)
{\sin kr\over kr}{4\pi r^{2}dr\over V}=
-{4\pi\delta\over Vk^{3}}
\left[-y\cos y+\sin y\right]^{kR(1-\sigma_{i})}_{0} &+&\nonumber\\
{4\pi A\over Vk^{3}}
\left[-y\cos y+\sin y\right]^{kR}_{kR(1-\sigma_{i})}+
{4\pi C\over Vk^{3}}
\left[-y\cos y+\sin y\right]^{kR(1+\sigma_{o})}_{kR} &-&\nonumber\\
{2\pi B\over Vk(k+\pi /\sigma_{i}R)^{2}}
\left[\left(y+{\pi\over\sigma_{i}}-{\pi\over 2}\right)
\sin y+{\cos y\over k+
\pi /\sigma_{i}R}\right]^{R(k+\pi /\sigma_{i}R)-
{\pi\over\sigma_{i}}+{\pi\over 2}}_{R(1-\sigma_{i})
(k+\pi /\sigma_{i}R)-{\pi\over\sigma_{i}}+{\pi\over 2}} &-&
\nonumber\\
{2\pi B\over Vk(-k+\pi /\sigma_{i}R)^{2}}
\left[\left(y+{\pi\over\sigma_{i}}-{\pi\over 2}\right)
\sin y+{\cos y\over -k+
\pi /\sigma_{i}R}\right]^{R(-k+\pi /\sigma_{i}R)-
{\pi\over\sigma_{i}}+{\pi\over 2}}_{R(1-\sigma_{i})
(-k+\pi /\sigma_{i}R)-{\pi\over\sigma_{i}}+{\pi\over 2}} &-&
\nonumber\\
{2\pi D\over Vk(k+\pi /\sigma_{o}R)^{2}}
\left[\left(y+{\pi\over\sigma_{o}}-{\pi\over 2}\right)
\sin y+{\cos y\over k+
\pi /\sigma_{o}R}\right]^{R(1+\sigma_{o})(k+\pi /\sigma_{o}R)-
{\pi\over\sigma_{o}}+{\pi\over 2}}_{R(k+\pi /\sigma_{i}R)-
{\pi\over\sigma_{o}}+{\pi\over 2}} &-&
\nonumber\\
{2\pi D\over Vk(-k+\pi /\sigma_{o}R)^{2}}
\left[\left(y+{\pi\over\sigma_{o}}-{\pi\over 2}\right)
\sin y+{\cos y\over -k+
\pi /\sigma_{o}R}\right]^{R(1+\sigma_{o})(-k+\pi /\sigma_{i}R)-
{\pi\over\sigma_{o}}+{\pi\over 2}}_{R(-k+\pi /\sigma_{o}R)-
{\pi\over\sigma_{o}}+{\pi\over 2}}.
\label{dk}
\end{eqnarray}
This model allows for the description of any compensated spherical
underdensity with the choice of parameters $\delta , R$ and 
$\sigma_{i},
\sigma_{o}$. Figure (1) shows the radial density contrast (top) 
and the corresponding Fourier transform (bottom) for different 
shapes. Note that the latter is strongly scale dependent:
it is characterized by a peak at $k\simeq 2\pi/2R$,
corresponding to the diameter of the bubble.
On smaller wave numbers, the spectrum falls as $k$,
while the oscillations at higher $k$ describe the compensating 
shell.
Left panels display the dependence on the shell width by maintaining
$\sigma_{i}=\sigma_{o}$; right panels display the analysis of
asymmetric shells $\sigma_{i}\ne\sigma_{o}$. The variation of 
width and shape of the compensating shell does not affect 
substantially the amplitude and the oscillating behaviour
of the Fourier transform. This makes the results of the next 
Sections 
almost independent of the exact shell shape; for this 
reason we put the emphasis on the more interesting physical 
observables $R$ and $\delta$, and
simply assume $\sigma =2\sigma_{i}=2\sigma_{o}=.3$ 
(the solid lines in Fig.(1)). 
Also, independent reasons to neglect the dependence on the shell's
shape arise from CMB damping of anisotropies on very small 
angular scales (Silk damping), to be described in the next Section. 
Finally, note that the perturbation
(and its CMB counterpart) scales linearly
with $\delta$, that has been factored out in the axis labels.
 
In the next Section we shall consider the perturbation 
described here in the
cosmic medium at reheating and we will evolve it until decoupling.

\section{From reheating to decoupling}

Here we want to compute in detail the
CMB perturbations induced by the
remnants of inflationary bubbles from reheating until decoupling, 
leaving to the next Section the analysis of the corresponding 
CMB anisotropies. 
We perform the computations in a standard
scenario (flat CDM, low baryon density ($\Omega_{b}=.06$),
$h$=.5). As in \cite{BAO} we adopt the approach developed in 
\cite{HS}, since it is applicable to any kind of scalar linear 
perturbation and very accurate ($5\div 10\%$). 
We report here only the basic and intuitive features of the method.

Let us consider the Fourier modes (\ref{dk}) as initial condition
in the matter distribution at reheating.
Let the scale factor be $a(\eta )$ and the time variable be 
the conformal time $\eta=\int_{0}^{t}d\tau /a(\tau )$,. 
Each mode $(\delta\rho/\rho )(k,\eta )$ evolves according to the 
linear theory of perturbation and its gravitational 
potential induces the corresponding time evolving perturbation 
$(\delta T/ T)(k,\eta )$ in the photon baryon plasma. 
At decoupling, the latter imprints anisotropies 
$(\delta T/ T)(\theta ,\phi )$ in the CMB.
Summarizing, the process involves the following steps:
\begin{equation}
{\delta\rho\over\rho}(k,0)\rightarrow{\delta T\over T}(k,0)
\rightarrow
{\rm evolution}\rightarrow{\delta T\over T}(k,\eta_{D})\rightarrow
{\delta T\over T}(\theta ,\phi )\ \ .
\label{m}
\end{equation}

The photon-baryon plasma contains zero-point
adiabatic $(\delta T/ T)_{0}(k,\eta )$ and pure velocity 
$V(k,\eta )$ perturbations; at decoupling the
Sachs-Wolfe effect (SW, $(\delta T/ T)_{SW}(k,\eta )$) given 
simply by the gravitational potential, must be added. 
In the tight coupling 
approximation, the equations driving the time evolution 
are (\cite{HS})
\begin{equation}\label{hs1}
\ddot{\left({\delta T\over T}\right)_{0}}+
{\dot{a}\over a}{{\cal R}\over 1+{\cal R}}
\dot{\left({\delta T\over T}\right)_{0}}+
{k^{2}\over 3(1+{\cal R} )}{\left({\delta T\over T}\right)_{0}}=
-\ddot{\Phi}-{\dot{{\cal R}}\over 1+{\cal R}}\dot{\Phi}-
{k^{2}\over 3} \Psi\ \ ,
\end{equation}
\begin{equation}\label{hs2}
V=-{3\over k}
\left[\dot{\left({\delta T\over T}\right)}_{0}+\dot{\Phi}\right]
\ \ ,\ \ \left({\delta T\over T}\right)_{0}(\eta =0)=-{\Psi (0)\over 2}
\ ,\ V(0)=0\ \ ,
\end{equation}
where dot denote derivative with respect to $\eta$.
The quantity ${\cal R}$ is $3\rho_{b}/4\rho_{\gamma}$ 
(the subscripts $b,\gamma$ refer to baryons and photons 
respectively), and 
$r_{s}(\eta )=\int_{0}^{\eta}d\eta '/\sqrt{3(1+{\cal R})}$
is the sound horizon at $\eta$. The gravitational potentials are
\begin{equation}\label{phi}
\Phi=(1+z){3H_{0}^{2}\over 2k^{2}}
\left({\delta\rho\over\rho}\right)(k,\eta )\simeq -\Psi\ \ ;
\end{equation}
the little difference between 
$\Phi$ and $\Psi$, due to the radiation components of the cosmic
fluids, is taken 
into account in the computations.
Below the Silk damping scale, the 
photon baryon plasma perturbations are damped by 
a factor $\exp [-k^{2}/k_{D}^{2}(\eta )]$, 
where the damping scale $k_{D}^{-1}$ is of the order of ten 
comoving Mpc at decoupling and soon there after grows to
the horizon scale. 
We simply put the spectrum (\ref{dk}) as the initial density 
perturbation; through linear theory, encoded for the CMB in 
(\ref{hs1}\ref{hs2}), we follow its evolution until decoupling.

The above procedure allows to perform the first three 
steps indicated in (\ref{m}); however, since the perturbations
here are non-Gaussian, the computation of the CMB angular 
anisotropies can not be done automatically as for Gaussian
perturbations, but
requires some more details, that we leave to the next Section. 
Note that the above procedure is general and does not depend 
on the particular initial perturbation spectrum; therefore the
CMB perturbation resulting from the evolution of any 
non-Gaussian linear feature from reheating may be calculated
from its Fourier transform using the above procedure. 
Also, because of linearity, the time 
evolution of our bubbles or any kind of non-Gaussian 
perturbations present in the cosmic fluid at reheating does
not depend on the presence of another (e.g. Gaussian)
component in the perturbation spectrum. This is of particular
interest for first order inflation models
which provide two independent perturbation mechanisms, 
(bubbles and Gaussian fluctuations from the slow rolling field in
the present case) as already mentioned in the Introduction.

Let us concentrate now on the CMB perturbations induced by bubbles.
We evolve the photon baryon perturbations until decoupling, when
we antitransform to get the relevants effects as a 
function of $r$. Figure (2) shows the adiabatic (top),
Doppler (middle) and SW effect (bottom). 
The bubbles under investigation are charecterized by a central 
density contrast $\delta$ at decoupling that we put on the y-axis 
units, and by $R=10,15,20{\it h}^{-1}\ {\rm Mpc}$; these radii are 
comparable with the size of the observed large
voids in the galaxy distribution (\cite{EPD1},1997). 
The most interesting feature of the results shown in 
Fig.(2) can be noted by looking at the curves and the scales: 
the SW effect strictly does not extend beyond the size of the 
perturbation, while the adiabatic and Doppler ones
are not vanishing until $60\div 80{\it h}^{-1}\ {\rm Mpc}$ 
from the center of the bubble, {\it independently}
of its radius. This is infact the amplitude of the sound
horizon of the photon-baryon plasma at decoupling. 
Just like a pebble in a pond, the initial small bubbly
perturbation is propagating beyond the 
initial scale, reaching the scale of the sound horizon at the time 
in which we are examinating it.
A sound wave with negative and positive crests is well evident 
at the sound horizon position, both in the adiabatic and Doppler effects. 
We have further investigated the motion of this wave for the case 
$R=20h^{-1}$ Mpc. Fig.(3) shows
in detail the wave at different times. Two physically
expected effects are immediately evident: 
the wave is travelling outward the bubble and simultaneously 
damped due to the Silk damping, 
that is becaming more and more active with time. 

Also note as the SW effect (and roughly the adiabatic one)
follows the known amplitude (\cite{P}) for linear perturbations 
$\delta T/T\simeq -\delta  (R/H^{-1})^{2}$. 
We have not found a significative dependence
on the shell's shape: for instance, by taking $\sigma^{'}=\sigma /2$ 
and repeating the same computations, the results change of a few 
percent; this is due to the Silk damping, the effect of which 
is to erase the small scale CMB anisotropies at decoupling;
in fact, the damping exponential 
mentioned above is $1/e$ at about $10h^{-1}$ Mpc in the present 
CDM model; thus, the damping is particularly effective on the small 
scales involved by the shell (a few comoving Mpc).   

As we will see in the next Section, the above results imply 
that the CMB anisotropies of an inflationary linear bubble 
consist in general in a few concentric rings of alternate 
colour (sign of $\delta T/T$) on the sub-degree angular scale.

\section{Anisotropies}

The perturbations in the photon-baryon plasma are photographed 
at last scattering and described by the CMB anisotropy field:
\begin{equation}\label{alm}
{\delta T\over T}(\theta,\phi )=
\sum_{lm}a_{lm}Y_{lm}(\theta,\phi )\ \ .
\end{equation}
Concerning the anisotropy 
amplitude, the physical information is contained in the correlation
function 
\begin{equation}
C(\theta )=
\left<\left({\delta T\over T}\right)_{\bf n}
\left({\delta T\over T}\right)_{\bf n^{'}}\right>_{{\bf nn^{'}}
=\cos\theta}=\sum_{l}{\sum_{m}|a_{lm}|^{2}\over 4\pi}
P_{l}(\cos\theta )\ \ ,
\label{cl}
\end{equation}
where $P_{l}$ are the Legendre polynomials and the average is taken
over all the sky directions ${\bf n}$ and ${\bf n'}$ separated 
by the
angle $\theta$. The coefficients $\sum_{m}|a_{lm}|^{2}$
are conventionally 
renamed as $(2l+1)C_{l}$ in a Gaussian theory; for the functional 
properties of the $P_{l}$, the $l-$th coefficient describes the CMB
anisotropies at the angular scale $\theta =\pi /l$ (see \cite{WSS} 
and references therein). The
coefficients $C_{l}$ of the expansion in Legendre polynomials 
do not determine uniquely the anisotropies, but only their
amplitude at a given angular scale specified by $l$; they 
completely neglect the phases of the $a_{lm}$ coefficients, 
and therefore describe the anisotropies only if the phases 
are randomly distributed (Gaussian perturbations). 
If on the contrary the anisotropies are the outcome of some 
coherent, i.e. non-Gaussian, structure,  
the $C_{l}$ are indicative of the amplitude only, but not of 
the coherence. This point is of particular interest in 
\cite{ABOII}, where the subject is the impact of a bubbly 
distribution on the $C_{l}$ spectrum.
 
Here we want to compute the CMB anisotropies (\ref{alm})
generated by remnants of inflationary bubbles.
On a particular direction ${\bf n}$, the anisotropy is given by
\begin{equation}
\left({\delta T\over T}\right)_{\bf n}=\int_{0}^{\eta_{o}}
\left({\delta T\over T}\right)_{\bf n}(\eta)P(\eta)d\eta\ \ ,
\label{dtt1}
\end{equation}
where $P(\eta )$
is the probability for a photon to be last scattered
between times $\eta$ and $\eta+d\eta$; 
$(\delta T/T)_{\bf n}(\eta )$
is the CMB perturbation field at causal distance $\eta_{0}-\eta$ 
on the direction ${\bf n}$. The perturbations examined here are
spatially localized: they have a center and vanish beyond
a sound horizon distance from it. Consequentely, in performing
the computation (\ref{dtt1}), we must specify the position
of the bubble's center with respect to the LSS.
We choose to orient the polar axis towards the center of the 
bubble, so that due to its spherical symmetry, the anisotropy 
field (\ref{alm}) is a function only of the polar angle $\theta$, 
that is the angle between the bubble center direction and 
the particular 
line of sight {\bf n}; the same angular choice was made in 
\cite{BAO}. The remaining degree of freedom regards
the distance of the bubble on the line of sight; its
perturbation on the CMB depends on the distance of the 
bubble's center
from the LSS (hereafter $d$) and vanishes of course if the 
perturbing zone does not intersects the LSS: to exemplify this 
dependence, 
we shall display in the sequel three interesting cases: 
$d=-R,0$ and $+R$. Let us consider
a photon last scattered at some $\eta$ inside the bubble 
and running towards the observation point on some 
direction $\theta$; the $\delta T/T$ that it carries depends 
on several variables: $\theta$, $d$ and $\eta$ uniquely 
determine the radial coordinate 
(radial with respect to the bubble's center of course) 
at which it decouples; 
we compute the adiabatic, SW and Doppler effects and
the overall CMB perturbation
\begin{equation}\label{dtt2}
{\delta T\over T}(\theta ,d ,\eta )=
\left({\delta T\over T}\right)_{0}+
\left({\delta T\over T}\right)_{SW}+
V\cdot\cos\alpha\ \ ,
\end{equation}
where $\alpha$ is the angle between the radial direction and the
line of sight, and of course accounts for the direction dependence
of the Doppler effect. The above expression means that for each 
$\eta$ we antitransform  $(\delta T/T)(k,\eta )$ as obtained
from the procedure of the previous Section. Note that this is
particularly accurate, since the behaviour of the
Silk damping scale (that suddenly after the decoupling 
grows up to the horizon scale) is taken into account.
Finally we get the CMB anisotropies simply 
averaging on all $\eta$'s, weighted 
with the last scattering probability $P(\eta )$ (\ref{dtt1}).
The integrated SW effect, due to the time change of the 
gravitational potential while the photon crosses the bubble,
is not taken into account since it is completely negligible,
as in the non-linear case (\cite{BAO}).

Figure (4) reports the CMB anisotropies from bubbles lying
on the LSS; the radii are the same as in Fig.(2). In each
panel, three different lines show the interesting variation of
the angular dependence of $\delta T/T$ with the relative positions
of the LSS with respect to the bubble's center: $d=-R,0,+R$
(respectively short dashed, solid and long dashed). Of course the
whole signal scales with $\delta$ and grossly the features of 
Fig.(2) are respected. Note the effects of the
variation of $d$: the negative strong spot near the center, 
coming mainly from the adiabatic effect (see also Fig.(2)), is
depressed if the bubble is displaced behind or beyond the LSS.

Finally in Figure (5) the CMB sky contains
only one bubble with 
$R=20{\it h}^{-1}\ {\rm Mpc}$ and $\delta =.005$ (top), 
$\delta =.01$ (bottom), lying on the
LSS, corresponding to the solid line on the bottom panel in
Fig.(4); in the same sky region we have added a purely 
Gaussian $\delta T/T$ from the standard CDM model. 
The spherical imprint provided by the bubble is well evident
in the bottom panel, but rather hidden in the top panel, depending
of course on $\delta$. Note however that both bubbles
become strongly non-linear at the present, and therefore are able to 
affect substantially the structure formation (\cite{OBAM}).
The signal shows clearly its marked non-Gaussianity:
the sound wave propagating outward and the central spot
generated by SW and adiabatic effect determine the appearance
of a few fascinating concentric rings of alternate colour on the
sub-degree angular scale. As the mean anisotropy amplitude in Fig.(5)
is not different from the ordinary Gaussian case,
the distinctive imprints of these structures
on the CMB are to sought not as much in the anisotropy 
amplitude (\ref{cl}), but 
rather in the phase correlation encoded in higher order 
statistics. 

The above results have to be extended to a
realistic distribution of bubbles. 
The resulting maps will be compared with the
present and next high resolution CMB observations. We leave
these analysis to forthcoming works, beginning with 
the impact on the $C_{l}$ (\ref{cl}), in \cite{ABOII}.

\section{Conclusions}

The occurrence of first order
phase transitions in the early universe is largely predicted, and
its indirect detection through the present observable traces would
be of crucial importance for our understanding of fundamental
physics. If one or more of such processes 
occurs well before the end of inflation, 
the nucleated bubbles are stretched to large scales; this has very 
important consequences on the observability of these very high 
energy physics phenomena. Here we predict the traces left by 
inflationary bubbles on the CMB.
To achieve this aim, we first model analytically a bubble. 
We use the general linear theory
to get the Fourier transform of the corresponding 
perturbation in the photon baryon plasma and to evolve it until 
decoupling. Then, by considering a CMB photon decoupled at some
point inside the bubble, we antitransform to get the $\delta T/T$ 
carried by the photon; by repeating the above computation for 
each direction and for each
decoupling point weighted with the last scattering probability
function, we get the whole CMB anisotropy produced by 
an inflationary bubble. 

We find that the CMB perturbation propagates
on the scale of the sound horizon at decoupling 
($\simeq 1^{o}$ in the sky), and appears
as a fascinating series of concentric 
rings of alternate color (different sign of $\delta T/T$),
a general feature that remains true as the bubble is moved back and
forth with respect to the last scattering surface. 
Therefore, in comparison with ordinary anisotropies from
scale invariant perturbation spectra, 
the present signal presents a marked
non-Gaussianity. The overall mean amplitude of the signal 
from a bubble with radius $R$ and central density contrast 
$\delta\ll 1$
is of the order of $\delta (R/H^{-1})^{2}$, as expected for 
linear perturbations. Therefore, bubbles with comoving radii 
corresponding to the size of the observed 
large scale structures (tens of Mpc), and characterized
by a central density contrast $\delta\simeq 10^{-3}$, induce
a level of anisotropy to be interestingly compared with
the observed ones. 

If bubbles are produced during the inflationary era, 
the resulting fluctuation spectrum is a 
superposition of the standard Gaussian fluctuations from the 
slow rolling inflaton and the scale dependent non-Gaussian ones from
the bubbles. The results of the present paper will be used
to make quantitative predictions of the imprint of these 
inflationary models on the CMB. 
Since the angular scale involved by the
bubbles are well below $1^{o}$, our computations shall
be compared with the high resolution CMB maps provided by
the next MAP and Planck experiments.

\acknowledgments

I wish to thank Luca Amendola, Pierluigi Fortini 
and Franco Occhionero for their 
collaboration and warm encouragement.

\figcaption{Radial density profiles (top panels)
and their Fourier transforms (bottom panels). 
The overall properties of the k-spectrum are almost the same
in any of the case shown. Examples
of symmetric shells ($\sigma_{i}=\sigma_{o}=\sigma /2$)
are on the left: $\sigma =.2$ (short dashed line),
$\sigma =.3$ (solid line) and $\sigma=.5$ (long dashed line)
Cases of asymmetric shells are on the right: $\sigma_{i} =.2$,
$\sigma_{o} =.1$ (short dashed line),
$\sigma_{i}=\sigma_{o}=.15$ (solid line) and
$\sigma_{i}=.1$, $\sigma_{o} =.2$ (long dashed line).}
\figcaption{CMB perturbations at decoupling
induced by an inflationary bubble: adiabatic effect in 
the top, Doppler in the middle, and Sachs-Wolfe in 
the bottom panel.
The different lines refer to different radii as indicated.
The Sachs-Wolfe effect does not exceed the bubble size,
while the other effects generally extend on the sound horizon
for the photon baryon plasma at decoupling.}
\figcaption{The outward traveling waves in the adiabatic (top) and
Doppler (bottom) effects for a bubble with
$R=20h^{-1}$ Mpc at different redshifts. 
The reduction of the amplitude is due to 
the Silk damping, 
that is becaming more and more active with time.}
\figcaption{CMB anisotropies induced by bubbles of interesting
size near the LSS in different positions: $d=-R$
(short dashed line), $d=0$ (solid line) and $d=+R$
(long dashed line). The signal scales linearly with
the central density contrast and involves angular scales
up to $1^{o}$.}
\figcaption{A portion of microwave sky containing 
a purely Gaussian CDM $\delta T/T$ field together with the 
signature
of an inflationary bubble with $R=20{\it h}^{-1}\ {\rm Mpc}$, 
$\delta =.005,\ .01$ 
(respectively top and bottom) and
sitting exactly on the LSS ($d=0$). Horizontal scale is in
acminutes. Although the spherical
imprint is well evident in the bottom panel but 
only hidden in the top one, both bubbles
are non-linear by the present and capable to affect substantially the
large scale structure.}

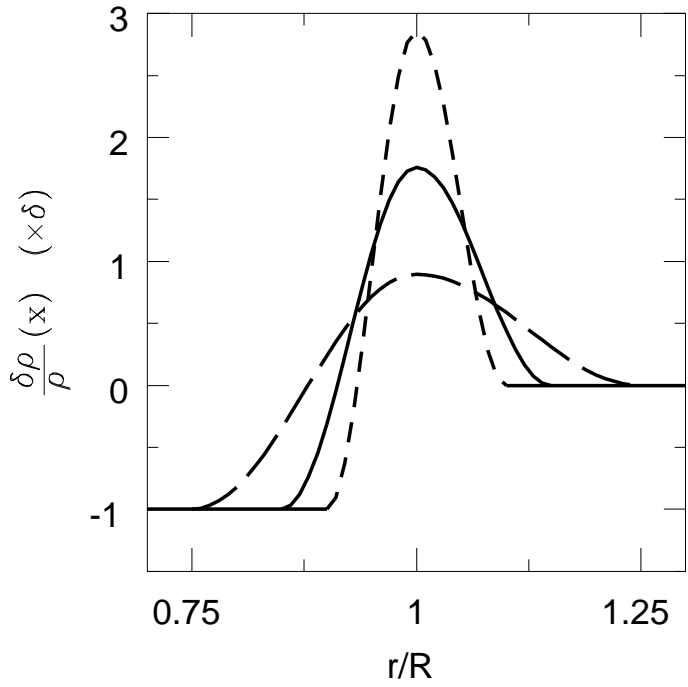
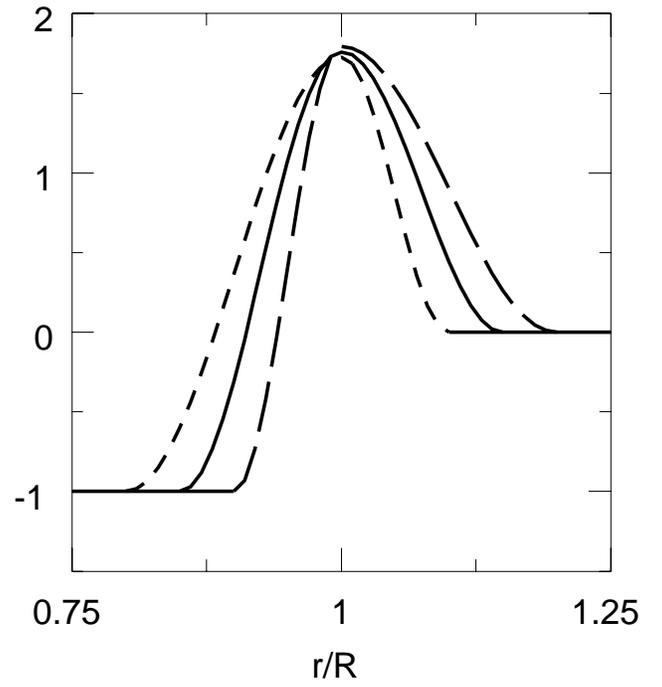
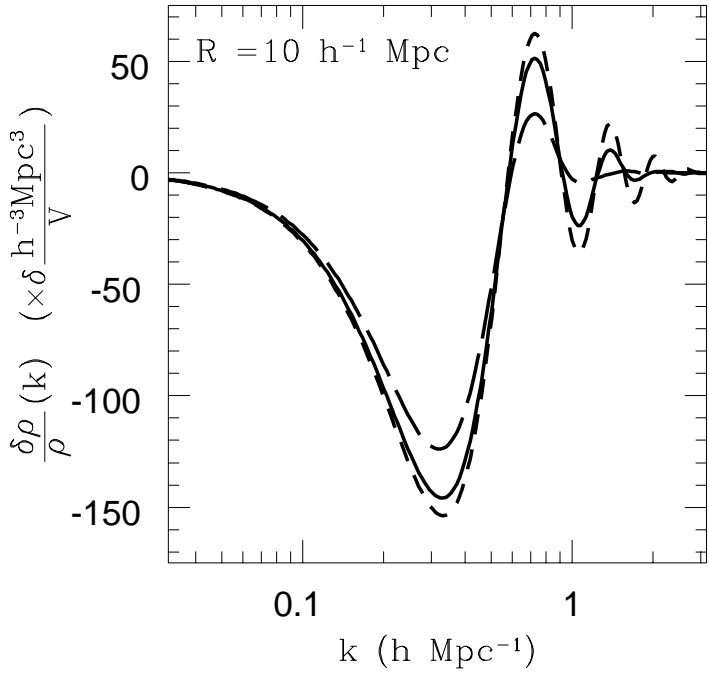
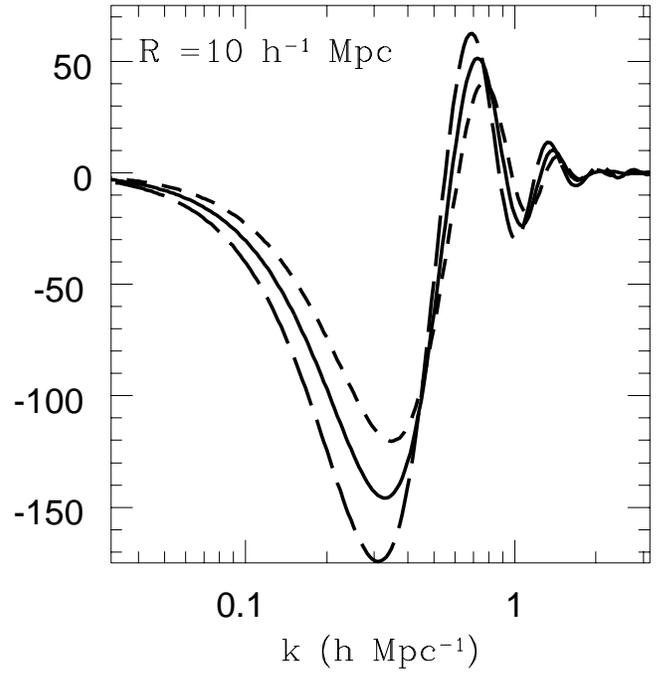
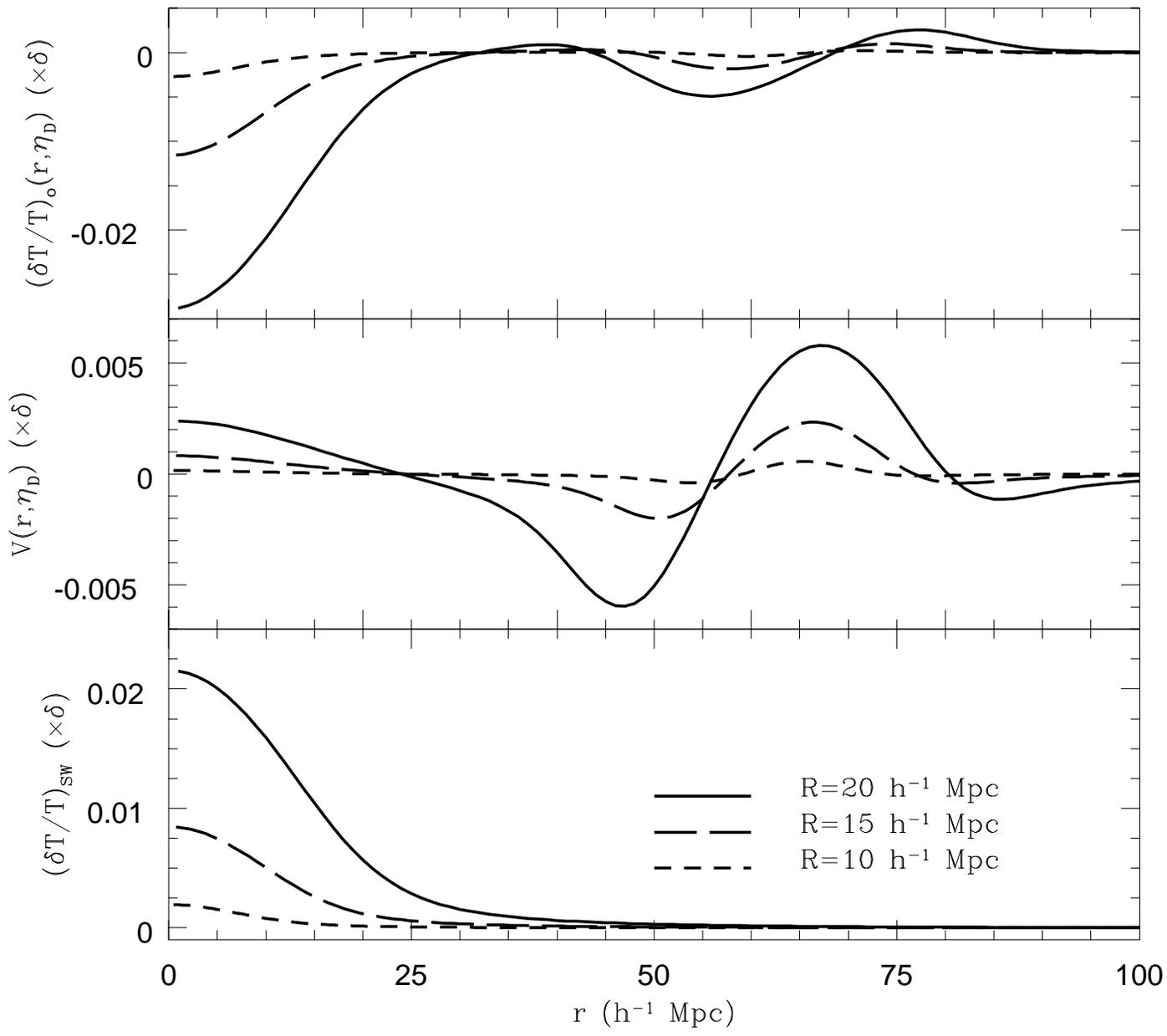
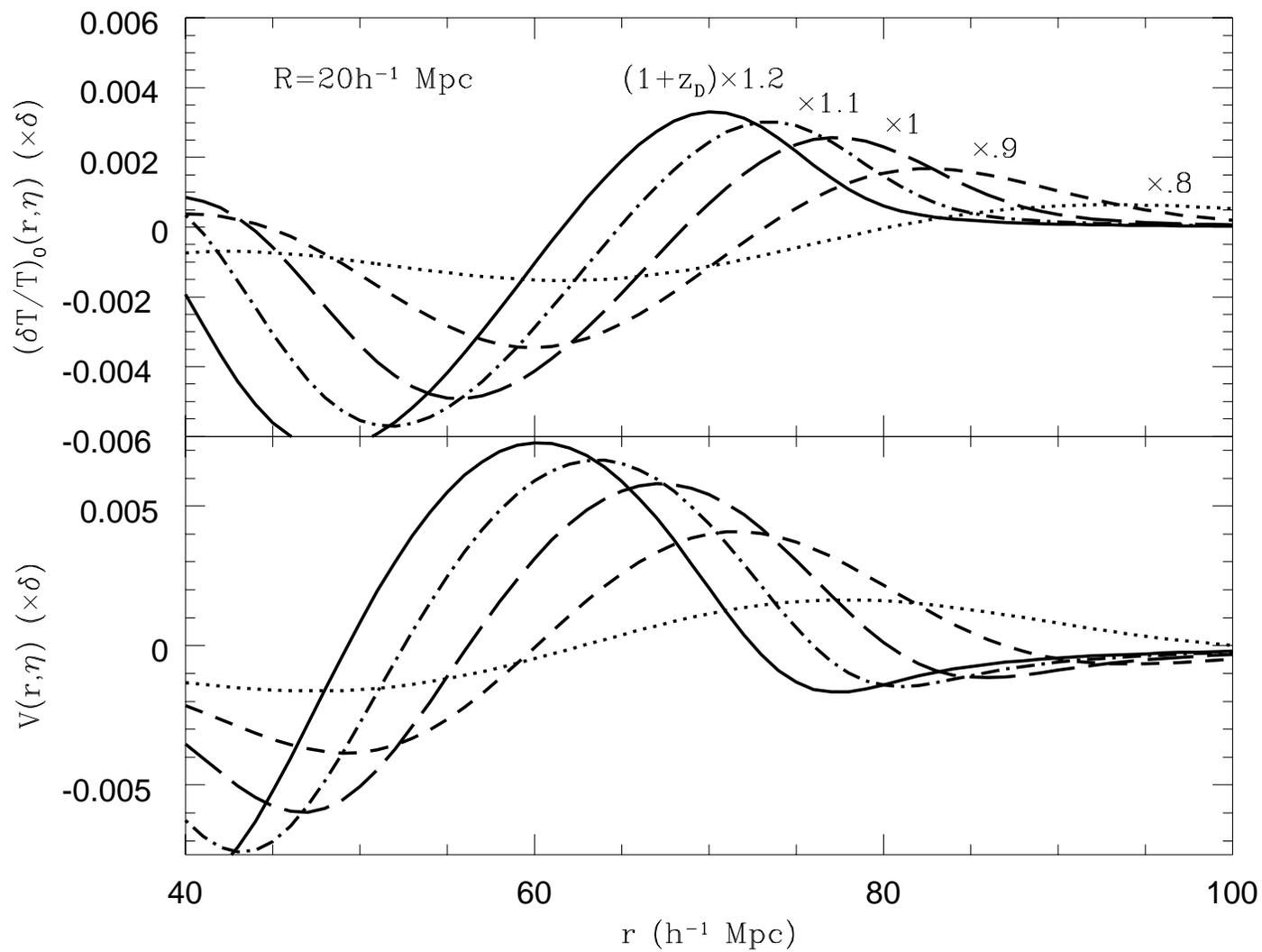
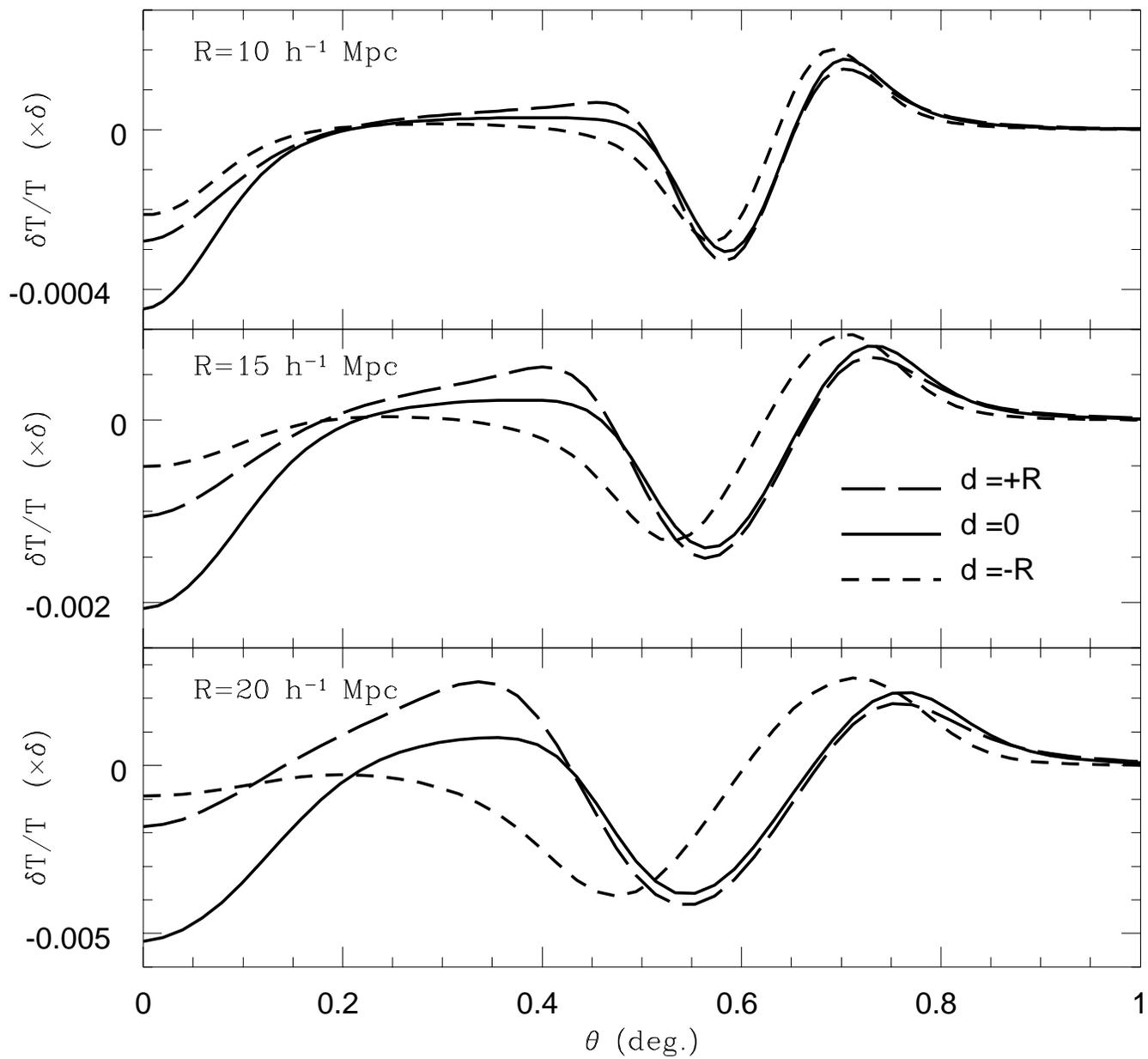
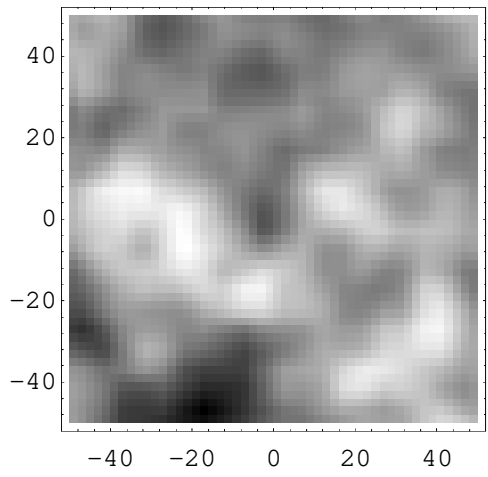
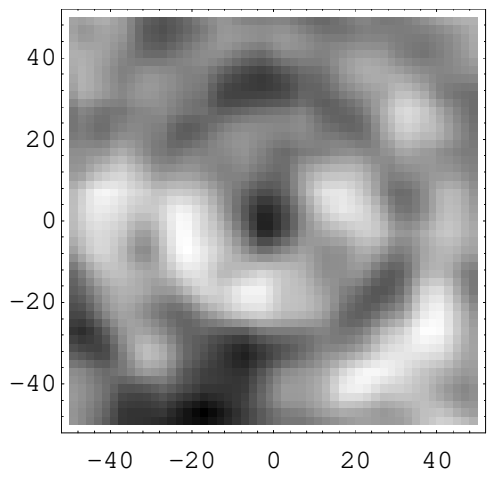
\end{document}